\documentclass[a4paper, 12pt]{scrartcl}
\usepackage{graphicx}
\usepackage{hyperref}
\hypersetup{
    colorlinks=true,
    citecolor={red},
    urlcolor={magenta},
    linkcolor={blue},
}
\usepackage{amsmath, amssymb, amsfonts, amsthm}

\graphicspath{{./img/}}

\usepackage{biblatex}
\title{How oscillations in SIRS epidemic models are affected by
the distribution of immunity times}
\author{Daniel Henrik Nevermann and Claudius Gros}
\date{}

\begin{document}
\maketitle
\begin{abstract}
Models for resident infectious diseases, like the SIRS model,
may settle into an endemic state with constant numbers of
susceptible ($S$), infected ($I$) and recovered ($R$) individuals,
where recovered individuals attain a temporary immunity to reinfection.
For many infectious pathogens, infection dynamics may also
show periodic outbreaks corresponding to a limit cycle in phase space.
One way to reproduce oscillations in SIRS models is to include
a non-exponential dwell-time distribution in the recovered state.
Here, we study a SIRS model with a step-function-like kernel 
for the immunity time, mapping out the model's full phase 
diagram. Using the kernel series framework, we are
able to identify the onset of periodic outbreaks when 
successively broadening the step-width. We further investigate 
the shape of the outbreaks, finding that broader steps cause more 
sinusoidal oscillations while more uniform immunity time 
distributions are related to sharper outbreaks occurring after
extended periods of low infection activity.

Our main results concern recovery distributions characterized
by a single dominant timescale. We also consider recovery
distributions with two timescales, which may be observed
when two or more distinct recovery processes co-exist. 
Surprisingly, two qualitatively different limit cycles are
found to be stable in this case, with only one of the two
limit cycles emerging via a standard supercritical Hopf 
bifurcation.

\bigskip
\end{abstract}

\section{Introduction}
The COVID-19 pandemic showed that it is important
to understand infection dynamics from two distinct
viewpoints. Firstly it is important to examine and
understand a given active outbreak, secondly a
thorough understanding of the general theory is
necessary, with the latter being the topic here.
Today, the Corona virus resides in an endemic state 
that is periodically disrupted by seasonal fluctuations or
sudden outbreaks followed by periods of less viral activity,
a behavior that is characteristic also
for other resident infectious diseases, like influenza, 
measles and pertussis \cite{martinez2018calendar}.

Synchronization phenomena in the form of oscillatory 
infection dynamics may be caused by a number of different 
interfering factors. External drivers include
seasonality \cite{liu2021role} or behavioral changes, 
potentially imposed by disease prevention measures 
\cite{di2020impact}. Other factors can be the nature
of the network of contacts \cite{li2014analysis}, travel
\cite{chinazzi2020effect} and even stochastic effects
\cite{aparicio2001sustained}. Taking a modeling point of 
view, the present study aims to improve the understanding 
of the origins of oscillations as arising from an
endogenous cause, namely from the distribution of immunity 
times when people are recovering from a disease.

Individuals in populations exposed to infective
pathogens typically pass through different stages, 
usually modeled via population subclasses. 
Non-infected individuals remain in the susceptible 
class $S$ until a contagion event happens, which 
moves the individual into the infected class $I$. 
Recovering from the disease, an individual usually
builds-up a temporary immunity, which puts the
individual into the class of recovered 
$R$.\footnote{Individuals in the $R$ class by be 
considered to be either `recovered', or `recovering',
depending on the terminology used. Note that infected (ill) 
individuals are considered to be immune too.}
Once immunity decayed, the individual returns 
to be susceptible, which completes the cycle. 
The standard SIRS model has the form
\begin{equation}
\label{eq:continuous-sirs-resolved}
    \begin{aligned}
        \dot{I} &= \beta I S - \rho I \\
        \dot{R} &= \rho I - \gamma R
    \end{aligned}, \quad\qquad 1 = S + I + R,
\end{equation}
where $\beta$ is the infection rate per individual, 
$\rho$ is the recovery rate and $\gamma$ is the 
immunity-fade rate. This formulation assumes 
exponential decay from a given to a subsequent 
compartment, where the inverse of the 
decay rates define characteristic dwell times. 
While SIRS models of the form 
\eqref{eq:continuous-sirs-resolved} capture damped 
oscillations towards a stationary state, the
endemic state, oscillatory states and/or states
with periodic outbreaks cannot be modeled.

Recently, several modifications to the standard SIRS model 
have been proposed, with some versions able to produce 
sustained oscillations, typically when time delays are
present. In \cite{Wagner_etal_2023} the authors try 
to incorporate human behavior by including mitigation
effects driven by perceived hazards, as characterized 
through past infection numbers. Sustained oscillations 
are obtained further-on when including time delays 
in the dwell-time distribution in the 
recovered compartment. Previous works considering 
non-exponential immunity time distributions 
\cite{Bestehorn_etal_2022, Gonçalves_Abramson_Gomes_2011} find
that oscillations appear depending on the specific choice of 
the delay kernel. If the kernel is given in terms of 
Erlang distributions,\footnote{Erlang distributions are of the form
$\lambda^n x^{n-1}\,\mathrm{e}^{-\lambda x}/(n-1)!$, 
with positive support, $x\ge0$. The parameters are 
$\lambda>0$ (real) and $n>1$ (integer).}
the kernel series framework
\cite{Nevermann_Gros_2023} can be applied to 
rewrite the time-delay model in terms of an 
extended set of ordinary differential equations. 
Similar approaches have been considered in the past, 
see \cite{Hurtado_Kirosingh_2019}. Alternatively,
the effect of non-trivial dwell-time distributions  
in the infected compartment has been examined, e.g.\ 
as in \cite{Greenhalgh_Rozins_2021}.

In this contribution, we consider models with
arbitrary immunity-time distributions,
compare \cite{Hethcote_1981}. Of special interest
will be immunity kernels having the form of 
a step function, which we denote block
delay kernels, and in particular broadened step functions,
here denoted soft steps, that allow to
interpolate smoothly between 
the classical SIRS model 
\eqref{eq:continuous-sirs-resolved},
for which immunity is distributed exponentially,
and the situation where immunity is identical
for all individuals. In the latter case, a mapping
to the corresponding discrete-time model is
possible. We find that stable limit cycles
describing non-uniform infectious dynamics 
emerge via a supercritical Hopf bifurcation
\cite{gros2024complex},
when either increasing average immunity times
and/or decreasing the width of the corresponding
smooth step.

The paper is organized as follows. In Sec.~\ref{sec:methods}
we discuss the SIRS model used in our
analyses, featuring a soft step delay
distribution for the time individuals retain immunity.
Before presenting the model we start with an outline
of the kernel series framework and
a SIRS model with general immunity time kernel. Afterwards,
in Sec.~\ref{sec:results}, we present our results for the
onset and the shape of periodic outbreaks when varying the
step width. Finally, we discuss our results in
Sec.~\ref{sec:discussion}. In the supplementary material,
Sec.~\ref{sec:sm-two-step-kernel}, we further present
results for a model with two timescales in the recovery distribution.

\section{Methods}
\label{sec:methods}
By applying the kernel series framework to a generalized SIRS 
model with arbitrary immunity time kernel we aim to systematically 
study the effects of broadened memory kernels. In particular, we 
will focus on step-function-like kernels we denote block delay 
kernels. We start with a brief review of the kernel 
series framework. For a comprehensive discussion see \cite{Nevermann_Gros_2023},
or literature on the linear chain trick, e.g.\ 
\cite{Hurtado_Kirosingh_2019,Hurtado_Richards_2020}. Afterwards, we
present a generalized SIRS model and specify to block delay kernels
and soft steps.

\subsection{Kernel series framework}
Using the kernel series framework, delay differential 
equations with distributed time delays can be translated 
to a higher dimensional system of ordinary
differential equations, granted the time 
delay kernel is given as a superposition of 
Erlang kernels. The latter is however not a 
limitation, given that a general delay kernel 
may be expanded in terms of Erlang functions
\cite{Bergé_etal_2019,Nevermann_Gros_2023}.
Consider exemplary the simple case of a one-dimensional 
system with a time delay distributed according to a 
single Erlang kernel, denoted here as
$K_m^{(N, T)}$,
\begin{equation}\label{eq:ksf-example-system}
    \dot{x}(t) = F\big(x(t), x_T(t)\big), \qquad\quad
    x_T(t)=\int_{0}^{\infty} x(t-\tau) K_N^{(N, T)}(\tau)\:\mathrm{d}\tau~,
\end{equation}
where the Erlang distribution is defined as
\begin{equation}
\label{def_erlang}
    K_m^{(N, T)} = \frac{N^m\tau^{m-1} \mathrm{e}^{-N\tau/T}}{(m-1)!\,T^m},
    \quad\qquad m, N \in \mathbb{N}~.
\end{equation}
In \eqref{eq:ksf-example-system}, the flow is given by
a generic non-linear function $F=F(x,x_T)$, where $x_T=x_T(t)$
is the delay term. Using the kernel series framework, 
one can recast \eqref{eq:ksf-example-system} exactly
to the following $(N+1)$-dimensional system of ordinary 
differential equations:
\begin{equation}\begin{aligned}\label{eq:ksf-ode-system}
    \dot{x}(t) &= F(x(t), x_N(t)) \\
    \dot{x}_1(t) &= \frac{N}{T} (x(t) - x_1(t)) \\
    \dot{x}_m(t) &= \frac{N}{T} (x_{m-1}(t) - x_m(t)), 
    \qquad\quad m = 2, \dots, N~,
\end{aligned}\end{equation}
where we introduced a set of auxiliary variables
\begin{equation}\label{eq:x_m}
    x_m(t) = \int_{0}^{\infty} x(t - \tau) K_m^{(N, T)}(\tau) \:\mathrm{d}\tau, 
    \quad\qquad m = 1, \dots, N~.
\end{equation}

\begin{figure}[t]
    \centering
    \includegraphics[]{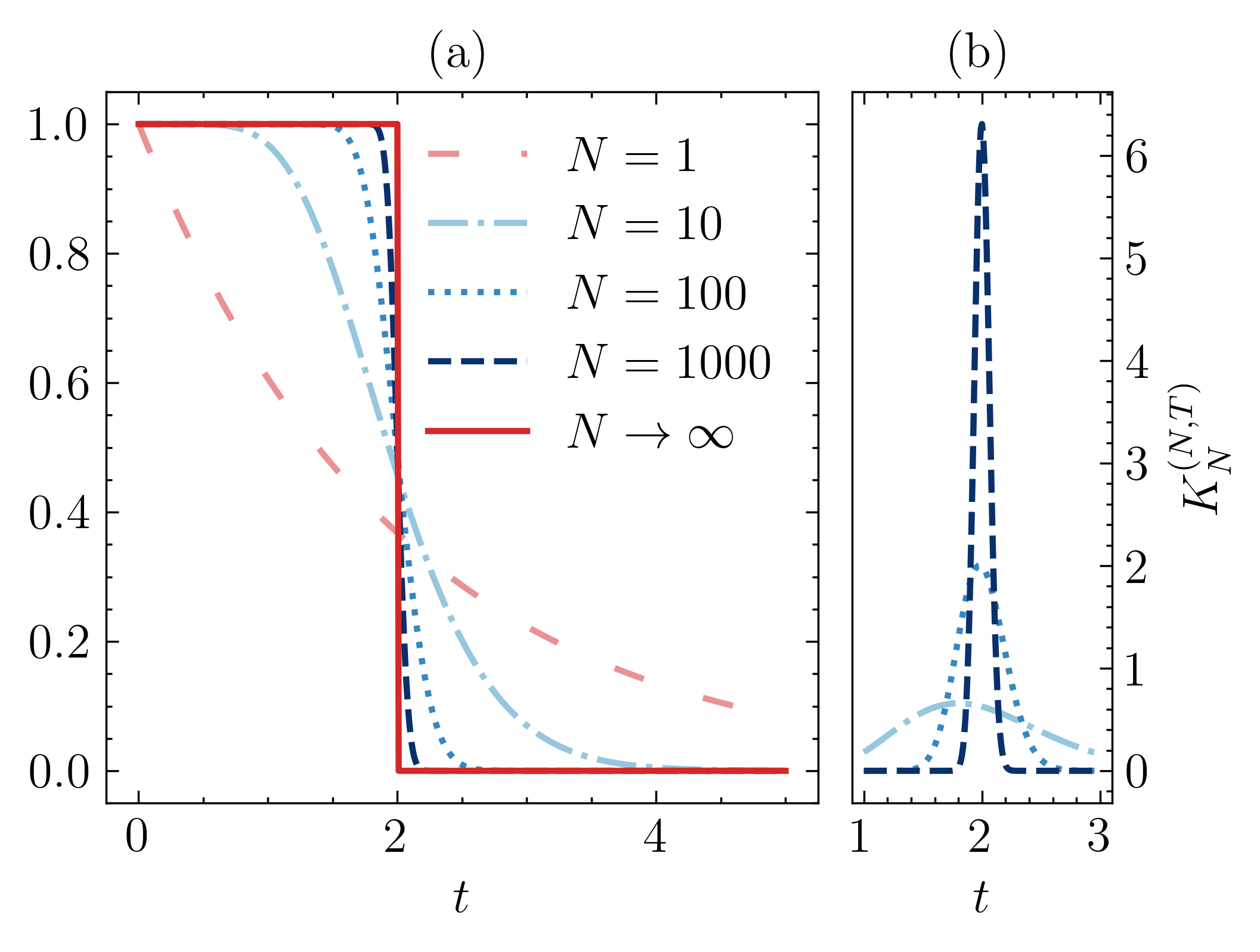}
    \caption{For $T=2$, an illustration of the step-function kernel
    defined in \eqref{eq:def-step-kernel} in \textbf{(a)}.
    Changing $N$, allows to interpolate between a block delay
    kernel ($N\to\infty$), presenting the limiting case of a
    discontinuous step function, and an exponential
    kernel ($N = 1$). \textbf{(b)} The step widths are determined
    by the standard derivative \eqref{d_theta_d_t}, given by 
    the widths of Erlang distribution $K_N^{(N,T)}$, as defined in
    \eqref{def_erlang}. Therefore, step widths are $\propto 1/\sqrt{N}$,
    see \eqref{eq:var-erlang-kernel}.}
    \label{fig:block-kernel}
\end{figure}

Since Erlang kernels converge to $\delta$-peaks,
\[
    \lim_{N\to\infty} K_N^{(N, T)}(t) = \delta(t-T)~,
\]
the distributed delay included in 
\eqref{eq:ksf-example-system} will converge 
to a fixed time delay, 
$x_T=x(t-T)$, in the limit $N\to\infty$, 
viz when the dimensionality of the corresponding 
ODE system \eqref{eq:ksf-ode-system} diverges. 
Of interest is also the variance of the Erlang kernel 
$K_N^{(N,T)}$, which is related to $N$ through
\begin{equation}\label{eq:var-erlang-kernel}
    \sigma^2 = \frac{T^2}{N}~.
\end{equation}
Thus, by varying $N$, the kernel series framework can be used as a handy
mathematical tool to investigate the response of a system to a broadened memory
kernel. An illustration is given in 
Fig.~\ref{fig:block-kernel}~(b).

\subsection{General immunity time kernel}
\label{sec:gen-imm-time-kernel}
In the standard SIRS model \eqref{eq:continuous-sirs-resolved}
the duration of immunity in a population is assumed
to follow an exponential distribution.
Generalizing to a generic
dwell-time distribution in the recovered compartment leads to
\begin{equation}\begin{aligned}\label{eq:general-sirs}
    \dot{I} &= \beta I (1 - I - R) - \rho I \\
    R &= \rho \int_0^\infty I(t-\tau) K(\tau)\:\mathrm{d}\tau
\end{aligned}\quad\qquad 1 = S + I + R~,
\end{equation}
including a distributed time delay with an arbitrary
delay kernel $K(t)$. As expected, the original SIRS model
\eqref{eq:continuous-sirs-resolved} is recovered when
using an exponential immunity time kernel
$K(t) = \gamma \exp(-\gamma t)$ in \eqref{eq:general-sirs}.

In order to apply the kernel series framework, 
one expands the kernel $K(t)$ in terms of Erlang kernels, 
which is generally possible, see
\cite{Bergé_etal_2019,Nevermann_Gros_2023}. 
For simplicity, we restrict ourselves here 
to the case that the expansion involves only 
Erlang kernels with fixed $N$ and $T$,
\begin{equation}\label{eq:K-ansatz}
    K(t) = \frac{T}{N}\sum_{m = 1}^{N}c_m K_m^{(N, T)}(t)~,
\end{equation}
where the prefactor is chosen such that $\lim_{t\to 0} K(t) = 1$. 
In analogy to \eqref{eq:x_m}, we define a series of auxiliary
compartments for recovering individuals,
\[
    R_m = \frac{\rho T}{N} \int_0^\infty I(t-\tau) 
    K_m^{(N,T)}(\tau)\:\mathrm{d}\tau, \quad\qquad m = 1, \dots, N~,
\]
Trivially, computing the $c_m$-weighted sum of the auxiliary compartments yields
the total recovered compartment
\[
    R = \sum_{m = 1}^{N} c_m R_m~.
\]
Inserting the ansatz \eqref{eq:K-ansatz} into \eqref{eq:general-sirs} gives
the SIRS model
\begin{equation}\begin{aligned}\label{eq:general-kernel-odes}
    \dot{I} &= \beta I \left(1- I - \sum_{m = 1}^N c_m R_m\right) - \rho I \\
    \dot{R}_1 &= \rho I - \frac{N}{T} R_1 \\
    \dot{R}_m &= \frac{N}{T} (R_{m-1} - R_m)
    \qquad\quad m=2,...,N~,
\end{aligned}\end{equation}
which constitutes the basis of our investigations.

\subsection{Smooth step kernels}
Specifying the expansion coefficients $c_m$ in
\eqref{eq:general-kernel-odes}, distinct immunity-time
kernels are generated. A basic choice is $c_m = 1$, 
which produces immunity distributions describing
broadened step functions, we here denote \emph{soft steps}
\begin{equation}\label{eq:def-step-kernel}
    K(t) = \Theta_N^{(T)}(t) = \frac{T}{N} \sum_{m = 1}^{N} K_m^{(N, T)}(t) \quad\xrightarrow{N\to\infty}\quad
    \begin{cases}1, & \text{for } 0\leq t\leq T \\ 0, & \text{else}\end{cases}~,
\end{equation}
where $T$ defines the location of the step and the parameter $N$ governs
the step width, smoothly transitioning between two extremes: a discontinuous
step function, we refer to as a \emph{block delay kernel}, as $N \to \infty$,
and an exponential distribution as $N \to 1$.
The kernels are illustrated in Fig.~\ref{fig:block-kernel}~(a).
From the derivative,
\begin{equation}
\label{d_theta_d_t}
    \frac{\mathrm{d}\Theta_N^{(T)}}{\mathrm{d}t} = -K_N^{(N, T)}(t) < 0~,
\end{equation}
one sees that the soft step kernels decrease strictly 
monotonic, and that the width of the step 
is given by the standard deviation of the highest order 
Erlang kernel in the series, which is $\propto 1/\sqrt{N}$, 
see \eqref{eq:var-erlang-kernel}.
The respective Erlang kernels are depicted 
in Fig.~\ref{fig:block-kernel}~(b).

For the above choice of the expansion coefficients, $c_m=1$, 
after application of the kernel series framework, we find that 
the corresponding SIRS model reads
\begin{equation}\begin{aligned}\label{eq:block-kernel-odes}
    \dot{I} &= \beta I \left(1- I - \sum_{m = 1}^N R_m\right) - \rho I \\
    \dot{R}_1 &= \rho I - \frac{N}{T} R_1 \\
    \dot{R}_m &= \frac{N}{T} (R_{m-1} - R_m)~,
\end{aligned}\end{equation}
which is derived from \eqref{eq:general-kernel-odes}. The SIRS model
\eqref{eq:block-kernel-odes} was previously studied in \cite{Hethcote_1981}. We
here enrich the studies on this model by numerically answering a similar
question previously posed in \cite{Gonçalves_Abramson_Gomes_2011}, namely the
onset of periodic outbreaks when broadening the step of a soft step kernel for
the time of immunity and furthermore studying the shape of the periodic outbreaks.

\section{Phase Diagram}
\label{sec:results}
Using the SIRS model \eqref{eq:block-kernel-odes} with a step-function-like kernel
we study the impact of soft step immunity kernels onto the onset and the shape
of periodic outbreaks, where we apply the kernel series framework and therein
the order parameter $N$ to interpolate between an exponential kernel ($N\to 1$)
and a block delay kernel ($N\to\infty$).

\subsection{Onset of periodic outbreaks}

The (non-oscillatory) endemic state of the model is found at the fixpoint, i.e.\
at the root of the flow equations in \eqref{eq:block-kernel-odes}. Solving the
system of equations yields
\[
    I_\mathrm{end} = \frac{\beta - \rho}{\beta (T \rho + 1)}~.
\]
As the width of the step function narrows, systematically controlled by
increasing the order parameter $N$, the system undergoes a supercritical
Hopf bifurcation, destabilizing the endemic state $I_\mathrm{end}$ and 
thereby giving rise to sustained oscillations. Hence, the onset of periodic outbreaks
coincides with the Hopf bifurcation point, which is defined as the root of the
maximal eigenvalue's real part, $\operatorname{Re}(\lambda_\mathrm{max})$, of
the system Jacobian evaluated at $I_\mathrm{end}$,
\[
    \mathbf{J}\big|_{I_\mathrm{end}} = \begin{pmatrix}
        \frac{\rho - \beta}{T \rho + 1} & \dots & &\\
        \rho & -\frac{N}{T} & &\\
        & \frac{N}{T} & - \frac{N}{T} & \\
        & & \ddots & \ddots
    \end{pmatrix}~.
\]

In the following, for our numerical studies, we use the fixed parameter
values $\beta = 2$ and $\rho = 1$
therefore assuming a basic reproduction number $R_0 = 2$, which is a realistic
figure for virus endemics such as influenza
\cite{Biggerstaff_Cauchemez_Reed_Gambhir_Finelli_2014}.

\begin{figure}[t]
    \centering
    \includegraphics[]{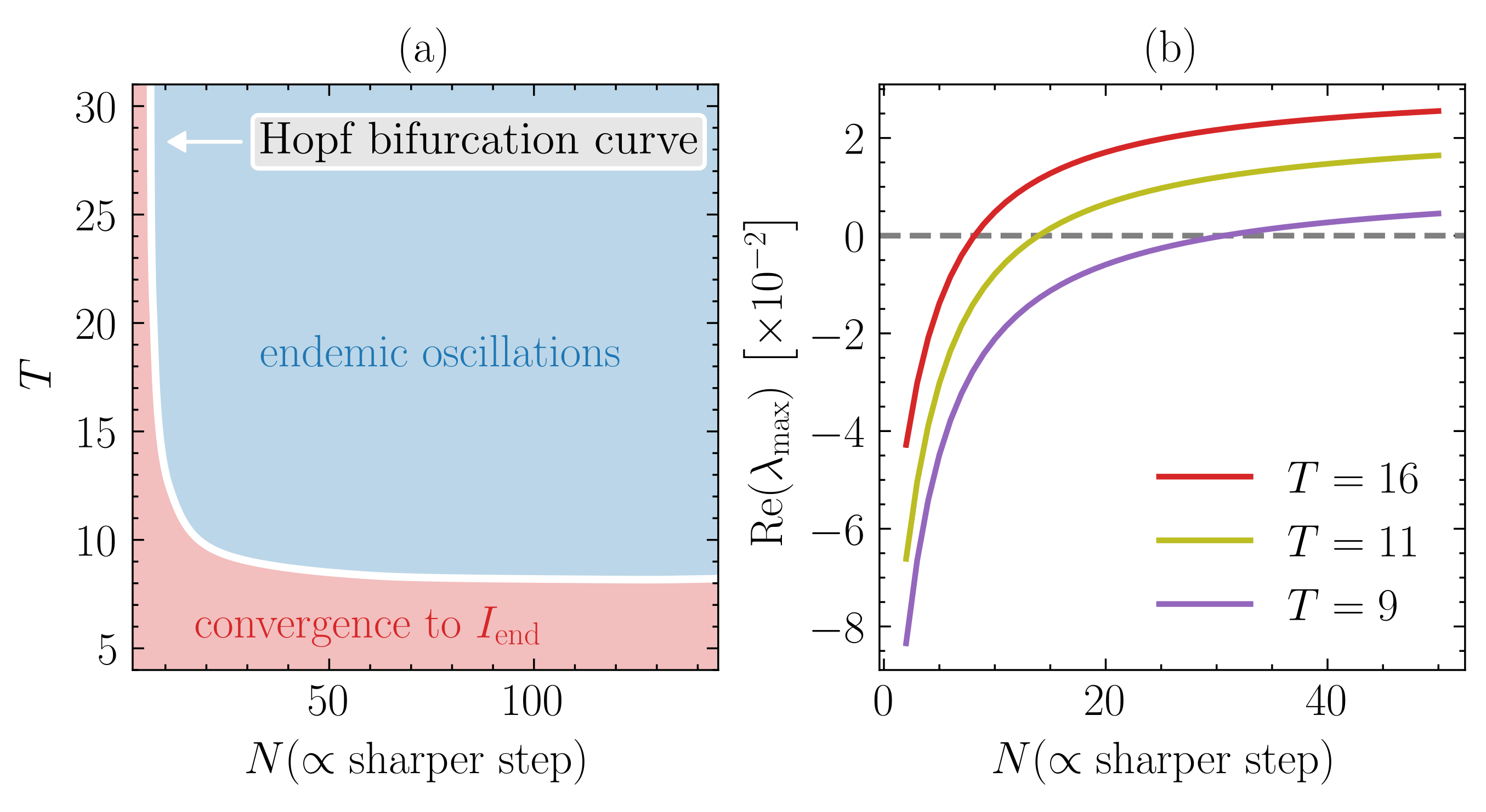}
    \caption{Onset of periodic outbreaks characterized through a Hopf
    bifurcation point, which occurs at the root of the maximal eigenvalue's 
    real part of the system Jacobian evaluated at $I_\mathrm{end}$,
    $\operatorname{Re}(\lambda_\mathrm{max})$. Positive values correspond to
    periodic outbreaks. In \textbf{(a)} we show the region of the parameter
    landscape $(N, T)$, where periodic outbreaks occur for $\beta = 2$ and $\rho
    = 1$ shaded in blue, while the system converges to the endemic state for
    parameters in the red-shaded region. Selected horizontal cross-sections
    for $T = 9, 11, 16$ are given in \textbf{(b)}.}
    \label{fig:eigenvalues}
\end{figure}

In Fig.~\ref{fig:eigenvalues} we show results for the onset of periodic
outbreaks.
Periodic outbreaks can only appear in a subset of the full parameter landscape
$(N, T)$, as indicated by the bifurcation diagram in
Fig.~\ref{fig:eigenvalues}~(a). The step width necessary to observe periodic outbreaks 
is strongly influenced by the immunity time, imposed through the parameter $T$ in the kernel.
For larger immunity times, oscillations already occur at relatively wide steps, compare
Fig.~\ref{fig:block-kernel}~(a). However, as the immunity time approaches the Hopf
bifurcation curve, much narrower step widths are necessary for
the emergence of endemic oscillations, see Fig.~\ref{fig:eigenvalues}~(b).

\subsection{Shape of periodic outbreaks}
\begin{figure}[t]
    \centering
    \includegraphics[]{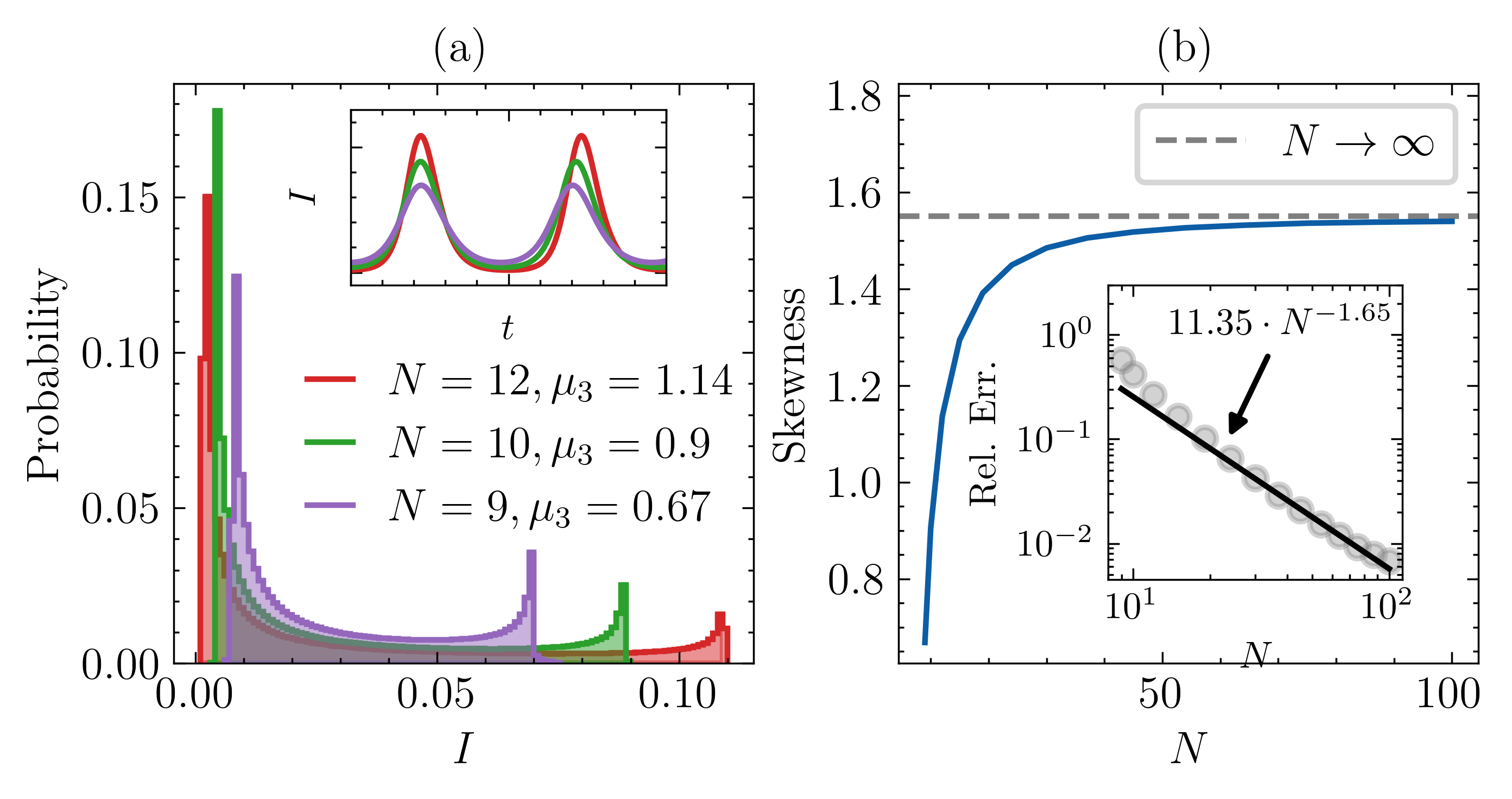}
    \caption{In \textbf{(a)}, probability densities associated to the time series 
    in the SIRS model \eqref{eq:block-kernel-odes} with $\beta = 2$ and $\rho = 
    1$ for different values of $N$, viz different step widths of the soft step
    kernel \eqref{eq:def-step-kernel}, along with the respective time series presented 
    in the inset. The 
    time constant in the kernels is kept at $T = 16$. Softer steps, i.e.\ 
    smaller figures of $N$, imply more sinusoidal periodic outbreaks and 
    therefore smaller positive skew. \textbf{(b)} The skewness converges to the 
    value found in the time-delay system at $N\to\infty$, where the relative 
    error to the system with (discontinuous) block delay kernel scales roughly as 
    a power-law $\propto N^{-1.65}$.}
    \label{fig:skewness}
\end{figure}

The impact of soft step delay kernels extends beyond the
appearance of periodic outbreaks. The shape of the periodic 
outbreaks also changes with the diversity of time scales
at which individuals loose immunity, given by the step with of
the soft step \eqref{eq:def-step-kernel}. Notably, we find that softer
steps render the periodic oscillations more
sinusoidal while a narrower step width induces sharp outbreaks 
separated by extended time spans where the disease is out 
of season. These tendencies can
be observed in Fig.~\ref{fig:skewness}.
As a compounded measure for the shape of the periodic outbreaks we employ the
skewness $\mu_3$ of the time series. For narrower step widths, the probability
density associated with the time series clearly shows increasingly positive
skew, see Fig.~\ref{fig:skewness}~(a). With increasing $N$, that is decreasing
step widths, the skewness quickly approaches an upper bound given by the
skewness of the time series in the limit $N \to \infty$. We show that the
relative deviation of the finite $N$ skewness from its upper bound scales
roughly as a power-law. Similar scaling of relative deviations for various
dynamical quantities in the kernel series framework was previously observed in
\cite{Nevermann_Gros_2023}.

\section{Discussion}
\label{sec:discussion}
Oscillations are an omnipresent feature of epidemic dynamics.
A comprehensive understanding is important for policymakers as 
well as epidemic modelers, in particular regarding preventive measures.
In the present work we investigated oscillations in time-delay SIRS 
models using a recently developed method, the kernel series 
framework \cite{Nevermann_Gros_2023}, which is applicable to 
arbitrary immunity time kernels, viz to any type of dynamics 
leading to the successive loss of immunity in the recovered 
population. A particular focus of our study concerns the relative
influence of two types of immunity dynamics, which may be lost
either progressively or rather abruptly, as described by 
immunity kernels having the form of a broadened step function, here
denoted soft steps.
We were able to identify the onset of oscillations with
respect to the softness of the step in the immunity time kernel 
and identify a relation between the step width with the shape of 
the periodic outbreaks.

The standard SIRS model \eqref{eq:continuous-sirs-resolved} 
takes a mean-field approach by including an exponentially 
distributed decay of immunity in the population, where
the inverse of the decay constant $1/\gamma$ defines the average 
time until an individual loses immunity. An alternative are
discontinuous delay kernels (with infinitely sharp steps),
which describes the case where every individual is immune 
for exactly the same period $T$. This limit, however, neglects 
population diversities. The kernel series framework 
\cite{Nevermann_Gros_2023} provides a convenient mathematical 
tool to probe the interpolation between the two limiting cases
by investigating the effect of smoothed block kernels.
Adding a second smoothed step to the block delay kernel (see
Sec.~\ref{sec:sm-two-step-kernel}) extends the model to a second
type of population diversity, namely in terms of two characteristic 
immunity times. Epidemiologically, one can imagine dividing the
population into subgroups based on factors such as pre-existing 
conditions (e.g., diabetes) or demographic factors (e.g., age or 
gender). That being said, we acknowledge that a realistic modelling 
of disease dynamics from which policy implications may be drawn would
require more complex immunity time kernels, potentially tailored to 
specific diseases. For this direction, further research is required.

Despite stylized assumptions for the immunity time kernels, the
reported implications of smoothed kernels onto shape and presence
of periodic outbreaks could be important in the crafting of infection
prevention plans. The presence of periodic outbreaks might require timed
countermeasures such as vaccination campaigns, lockdowns or behavioral changes.
Furthermore, different shapes of outbreaks might demand tailored prevention
strategies, where e.g.\ sudden severe outbreaks could require more directed
countermeasures than sinusoidal oscillations.
Our results highlight the importance of considering immunity time kernels
in epidemic forecasting.

Taking the timescale of the constant $T$ in the immunity time kernel, 
which is given in units of the mean dwell time in the infected state, 
$\langle T_\mathrm{infected}\rangle = 1/\rho$, allows to place 
oscillation periods and characteristic time delays into context. 
For a block delay kernel, we find that oscillations appear for 
immunity times $T \approx 8 \cdot \langle 
T_\mathrm{infected}\rangle$ (see Fig.~\ref{fig:eigenvalues} and
\cite{ripperger2020orthogonal}). For a two-step kernel the onset of 
oscillations occurs only at significantly larger values of $T$.

Contrary to the single-step model, we found multi-stability in 
the two-step model, see Sec.~\ref{sec:sm-two-step-kernel}, which 
implies a strong dependence on initial conditions. Similar 
multistabilities in time-delay SIRS models were found previously, 
e.g.\ in \cite{Bestehorn_etal_2022}. If the system resides close 
to its fixpoint it will either converge back to the fixpoint or 
show mild periodic outbreaks.
For different initial conditions, such as a patient zero in a 
susceptible population, the system asymptotically shows severe 
periodic outbreaks affecting up to $\approx 80\%$ of the population. 
The origins of endemic oscillations not induced through a Hopf 
bifurcation as well as potentially similar limit cycles in
models with more complex immunity kernels, such as three-step 
kernels, and their epidemiological interpretation are subject 
to future research. For policymakers, the potential existence
of more than one stable epidemic states would have sever 
implications. Applying a suitable chock therapy, like kick 
control \cite{sandor2018kick}, may be considered, with the
goal 'to kick' the system from an unfavorable attractor
into a comparatively more benign epidemic state.

%
\printbibliography
%

%
\appendix
%

\section{Two-step block delay kernel}
\label{sec:sm-two-step-kernel}
\begin{figure}[t]
    \centering
    \includegraphics{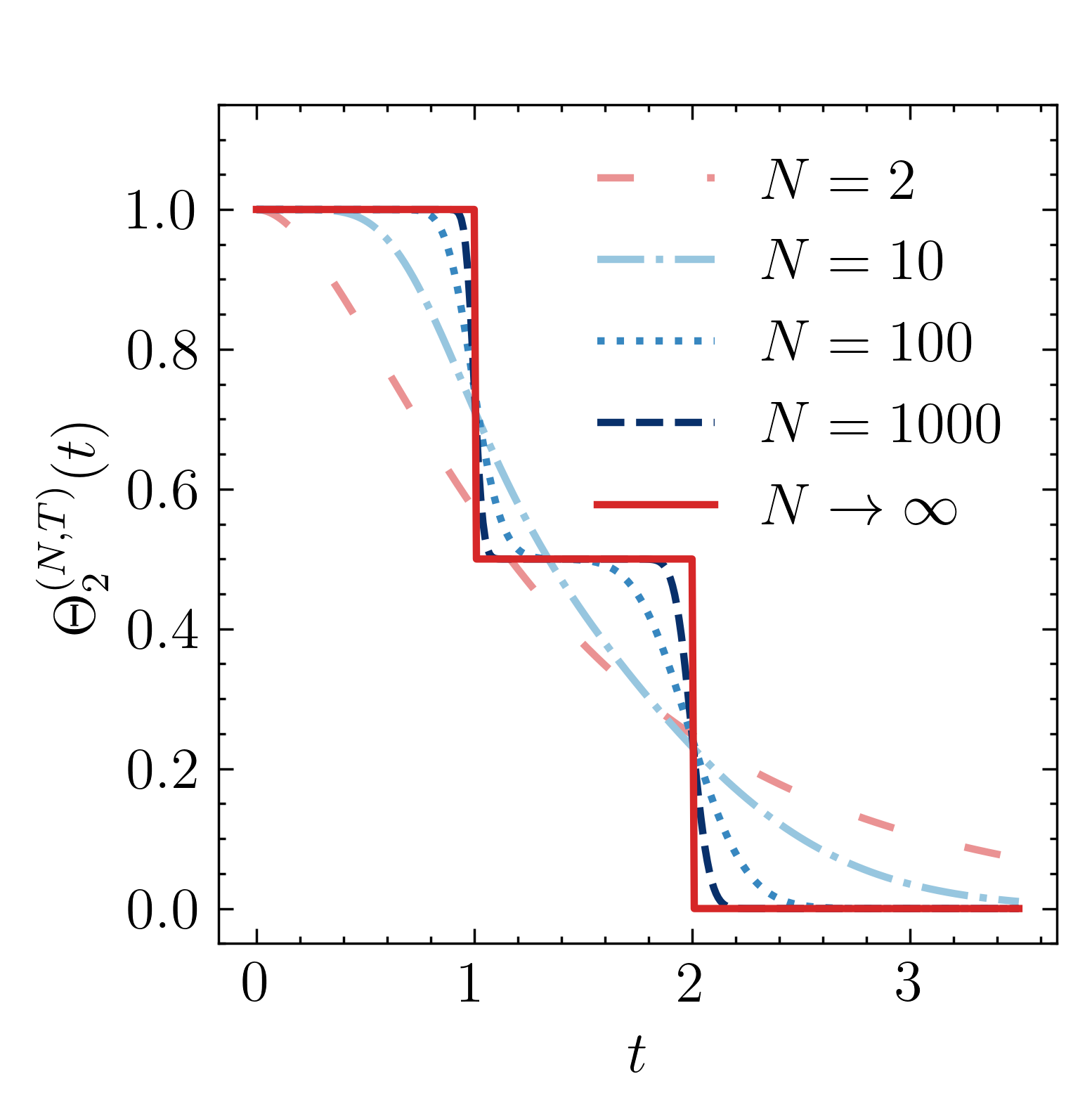}
    \caption{Two-step kernels $\tilde{\Theta}_N^{(T)}(t)$, as defined in
    \eqref{eq:two-step-kernel-def}, for $T = 2$ and different
    values of $N$. The double step results from a superposition of two
    step functions with time constants $T_1 = T/2$ and $T_2 = T$. For small $N = 2$,
    a quasi-exponential distribution is recovered.}
    \label{fig:two-step-kernel}
\end{figure}

The flexibility of the general immunity-time SIRS model introduced in
Sec.~\ref{sec:gen-imm-time-kernel} allows studying more complex immunity-time
kernels. The next logical step in the evolution of block delay kernels is the
introduction of a two-step kernel, that is
\begin{equation}\label{eq:two-step-kernel-def}
    K(t) = \tilde{\Theta}_N^{(T_1, T_2)}(t) = \frac{1}{2} \left(\Theta_N^{(T_1)}(t) + \Theta_N^{(T_2)}(t)\right)\quad\xrightarrow{N\to\infty} \quad\begin{cases}1, & \text{for } 0\leq t\leq T_1 \\ 0.5, & \text{for } T_1 < t \leq T_2 \\0, & \text{else}\end{cases}~,
\end{equation}
in which $N$ controls the width of the steps and $T_1, T_2$ control the loci of the
two steps, where in the following we choose $T_1 = T/2$ and $T_2 = T$.
The two-step kernel may also be written in terms of a series
expansion of Erlang kernels, thereby implicitly specifying expansion
coefficients $c_m$
\[
    \tilde{\Theta}_N^{(T/2, T)}(t) \equiv \tilde{\Theta}_N^{(T)} = \frac{T}{N} \left( \sum_{m = 1}^{N/2} K_m^{(N, T)}(t) + \frac{1}{2}\sum_{m = N/2+1}^{N} K_m^{(N, T)}(t) \right)~.
\]
For an illustration of the two-step kernel see Fig.~\ref{fig:two-step-kernel}.
The corresponding SIRS model then reads
\begin{equation}\begin{aligned}\label{eq:two-step-kernel-odes}
    \dot{I} &= \beta I \left(1- I - \sum_{m = 1}^{N/2} R_m - \frac{1}{2} \sum_{m = N/2+1}^{N} R_m\right) - \rho I \\
    \dot{R}_1 &= \rho I - \frac{N}{T} R_1 \\
    \dot{R}_m &= \frac{N}{T} (R_{m-1} - R_m)~.
\end{aligned}\end{equation}

\subsection{Endemic state and the onset of oscillations}
The non-oscillatory endemic state of the SIRS model
\eqref{eq:two-step-kernel-odes} is again found at the root of the flow, which
leads to
\[
    I_\mathrm{end} = \frac{\beta - \rho}{\beta (1 + \frac{3}{4} \rho T)}~.
\]
As the step widths increase, i.e.\ as the order parameter $N$ decreases,
$I_\mathrm{end}$ looses stability in a Hopf bifurcation which leads to
periodic oscillations. The onset of these endemic oscillations is again
found by computing the root of the maximal eigenvalue's real part of the Jacobian
\[
    \mathbf{J}\big|_{I_\mathrm{end}} = \begin{pmatrix}
        -\frac{4 (\beta - \rho)}{3 \rho T + 4} & \dots & -\frac{2 (\beta - \rho)}{3 \rho T + 4} & \dots \\
        \rho & -\frac{N}{T} & &\\
        & \frac{N}{T} & - \frac{N}{T} & \\
        & & \ddots & \ddots
    \end{pmatrix}~.
\]

\begin{figure}[t]
    \centering
    \includegraphics[width=0.88\textwidth]{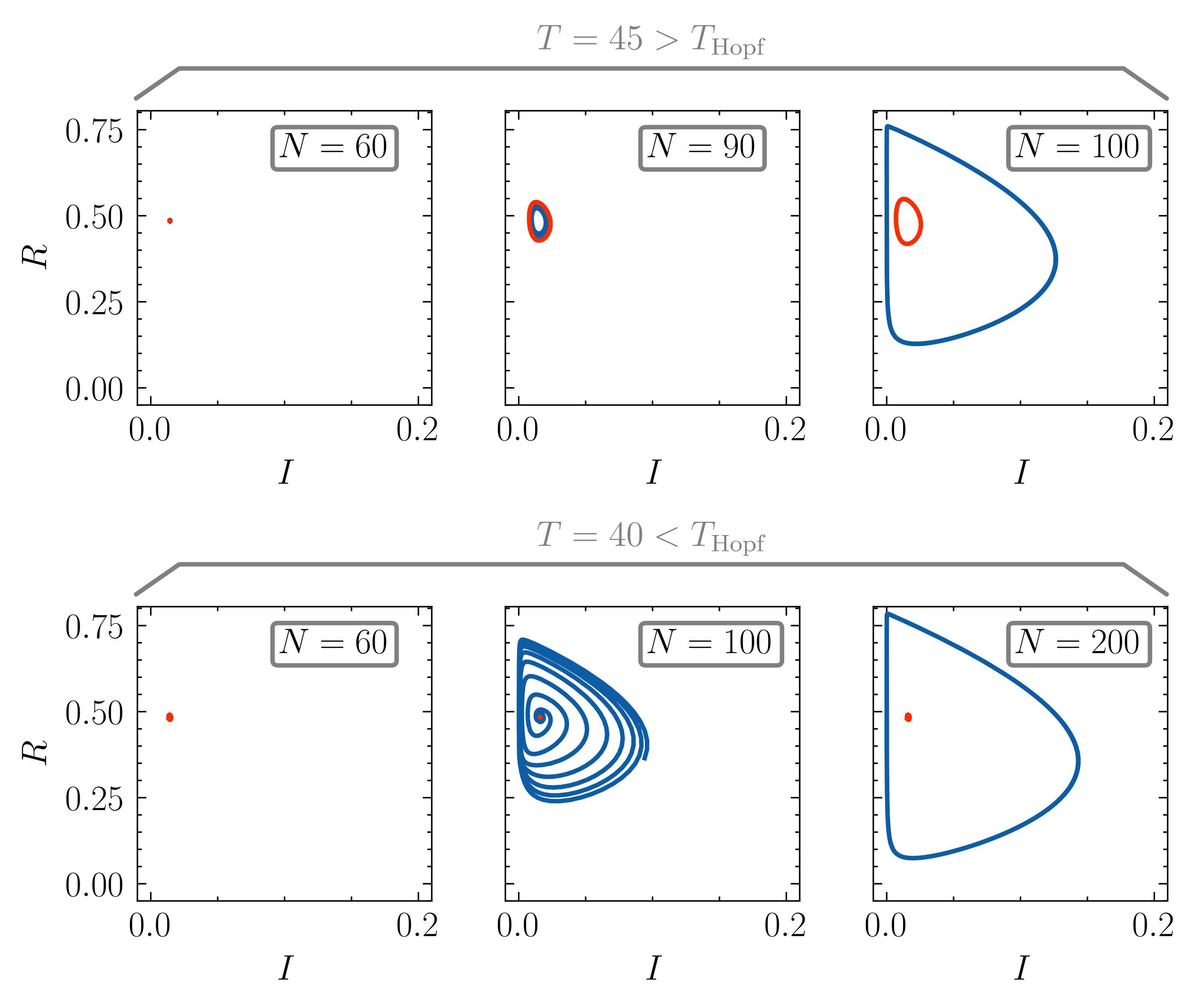}
    \caption{Phase space plots $(I, R)$ for the two-step kernel system
    \eqref{eq:two-step-kernel-odes} with $\beta = 2$ and $\rho = 1$. The
    top row shows the dynamics above the
    Hopf bifurcation curve, the bottom row shows the dynamics below the Hopf
    bifurcation curve. Red trajectories were sampled using an initial condition
    close to the fixpoint, whereas blue trajectories were sampled using initial
    conditions resembling a few initial infections in an otherwise
    pathogen-naive population. Multistability observed in the two-step model is
    shown in the right-most column.}
    \label{fig:two-step-phaseplots}
\end{figure}

Contrary to our expectations we could identify a second type of endemic
oscillations, which arises independently of the Hopf bifurcation induced
oscillations. For parameter values above the Hopf bifurcation curve we observe a
small stable limit cycle in the phase space $(I, R)$ triggered by the
destabilization of the endemic state but also a second larger stable limit
cycle, resembling severe periodic outbreaks with the period almost doubled compared to the
Hopf induced limit cycle, see Fig.~\ref{fig:two-step-phaseplots}. Importantly,
endemic oscillations persist below the Hopf bifurcation curve with the implication that
the onset of oscillations lies below the Hopf bifurcation.

\begin{figure}[t]
    \centering
    \includegraphics{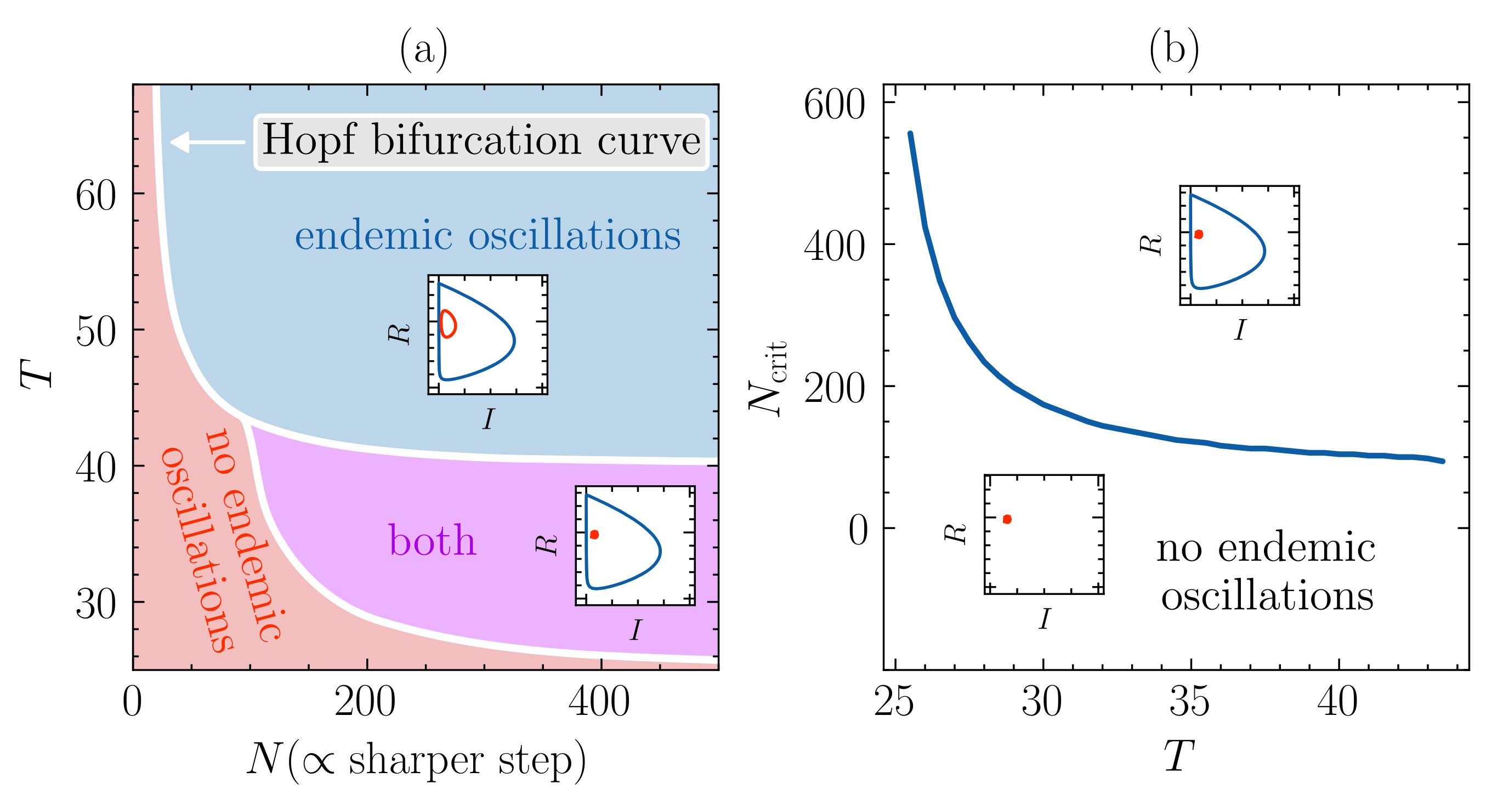}
    \caption{\textbf{(a)} Shows the full bifurcation diagram for the two-step
    kernel system \eqref{eq:two-step-kernel-odes} with $\beta = 2$ and $\rho =
    1$. Parameter regions shaded in red lead to a decay of oscillations to the
    constant endemic state. In the purple shaded region we observe
    multistability, where the system either decays to $I_\mathrm{end}$ or enters
    endemic oscillations, depending on initial conditions. In the blue shaded
    region the system shows endemic oscillations for all $(N, T)$. The specific
    shape of the oscillations depends on the initial conditions. In \textbf{(b)}
    we present the critical values of the order parameter $N_\textrm{crit}$ at
    which oscillations appear as a function of $T < T_\textrm{Hopf}$.}
    \label{fig:two-step-oscillation-onset.png}
\end{figure}

The full bifurcation diagram is shown in
Fig.~\ref{fig:two-step-oscillation-onset.png}~(a).
Compared to the model with single soft step kernel
\eqref{eq:block-kernel-odes} (see Fig.~\ref{fig:eigenvalues}), we find the locus
of the Hopf bifurcation greatly shifted upwards. Consequently,
destabilization of $I_\mathrm{end}$, as well as the onset of Hopf
bifurcation-induced oscillations, requires considerably larger immunity times $T
\gtrsim 40$.
Furthermore, the Hopf bifurcation curve is less steep which means that the
destabilization of $I_\mathrm{end}$ generally requires larger $N$ and thus
sharper steps. These results seem plausible as the two-step kernel is closer to an
exponential kernel than the block delay kernel, where in the exponential case no
stable oscillations occur at all and $I_\mathrm{end}$ remains stable.
On the other hand, unlike in the model \eqref{eq:block-kernel-odes} with a block
delay kernel, endemic oscillations are present below the Hopf bifurcation curve.
The asymptotic behavior of the system crucially depends on initial conditions.
If the system is close to the fixpoint, the fraction of infected individuals
will quickly converge to $I_\mathrm{end}$, whereas severe periodic outbreaks may
be observed even below the Hopf bifurcation if the system is further away from
the fixpoint and importantly also for initial conditions resembling a few
initial infections in an otherwise infection-naive population, see
Fig.~\ref{fig:two-step-phaseplots}.

For $T < T_\mathrm{Hopf}$ oscillations die out completely below a
critical value of the order parameter $N_\mathrm{crit}$. We find that
$N_\mathrm{crit}$ diverges almost exponentially with increasing distance to the
Hopf bifurcation curve, viz for shorter immunity times, oscillations only appear
for sharp steps, see Fig.~\ref{fig:two-step-oscillation-onset.png}~(b). We find
that in the limit $N\to\infty$ the potential for two limit cycles is retained. 

\end{document}